# Interfacing Ultraclean Graphene with Solid-State Devices

*Jean-Nicolas Longchamp\*, Conrad Escher, Hans-Werner Fink*

*Physics Institute, University of Zurich, Winterthurerstrasse 190, 8057 Zurich, Switzerland*

**\*Corresponding Author**

E-mail: longchamp@physik.uzh.ch

**ABSTRACT**

Interfacing graphene with solid-state devices and maintaining it free of contamination is a crucial step towards a functioning device, be it a semiconductor structure or any other device for technological applications. We take advantage of the catalytic properties of platinum metals to completely remove the polymer capping after the transfer of macroscopic graphene sheets to a solid-state device. For that purpose a platinum metal coated structure is brought in close proximity with the polymer capping. Subsequent annealing in air at a temperature between 175 and 350°C actuates a complete catalytic removal of the polymer. Finally, the platinum metal catalyst is removed revealing ultra clean graphene interfaced with an arbitrary device. Experiments to interface macroscopic graphene layers with oxidized silicon wafers demonstrate the general applicability of this approach. In repeating the procedure, multi-layer graphene sheets can also be produced. Direct evidence for the latter is provided by optical images of three overlapping graphene sheets. The exceeding level of cleanliness of the graphene is examined on the nanometer scale by means of low-energy electron transmission microscopy.

# INTRODUCTION

The physical and in particular the electronic properties of graphene depend to a large extent on its structure and cleanliness. Scattering of transport electrons at impurities is one of the major drawbacks in the use of graphene in electronic devices[1–3]. While the growth of single-layer graphene by means of chemical vapor deposition (CVD) is nowadays routinely possible[4,5], easily accessible and reliable techniques to transfer graphene to different substrates in a clean manner are still lacking. The common technique for the transfer of layers grown by means of CVD on a metallic substrate (usually nickel or copper) onto an arbitrary substrate is based on the use of a sacrificial polymer layer, usually polymethyl methacrylate (PMMA), spread or spin-coated over graphene[6–8]. The removal of the approximately 100 nanometer thick PMMA layer is a challenge and extensive efforts have been undertaken in the past few years to establish a reliable technique to retrieve pristine graphene without PMMA residues[1,2,6,9–11]. Typical chemical etchants for PMMA are acetone and chloroform. Unfortunately, wet chemical treatment of the polymer leads to graphene layers with a considerable amount of residues. Thermal annealing at temperatures in the range of 300-400°C in vacuum[11,12] or in an Ar/$H_2$ atmosphere[12] appears to forward the cleaning process, yet it does not lead to contaminant free graphene.

Recently, we have reported a method for preparing ultraclean freestanding graphene based on the removal of a sacrificial PMMA layer by the catalytic activity of Pt-metals[13,14]. However, after completion of the process, a thin layer of Pt-metal remains in close proximity to the ultraclean graphene obstructing its use for electronic applications.

Here, we present an expedient method for preparing ultraclean graphene on arbitrary substrates. Selectively, mono-, bi- and tri-layers of graphene have been prepared on SiO$_2$/Si substrates and inspected by optical interference microscopy. Moreover, freestanding graphene has been prepared over micron sized holes in a thin SiN membrane. The level of cleanliness of the freestanding graphene layers has been investigated by means of low-energy electron transmission microscopy[13,15–18].

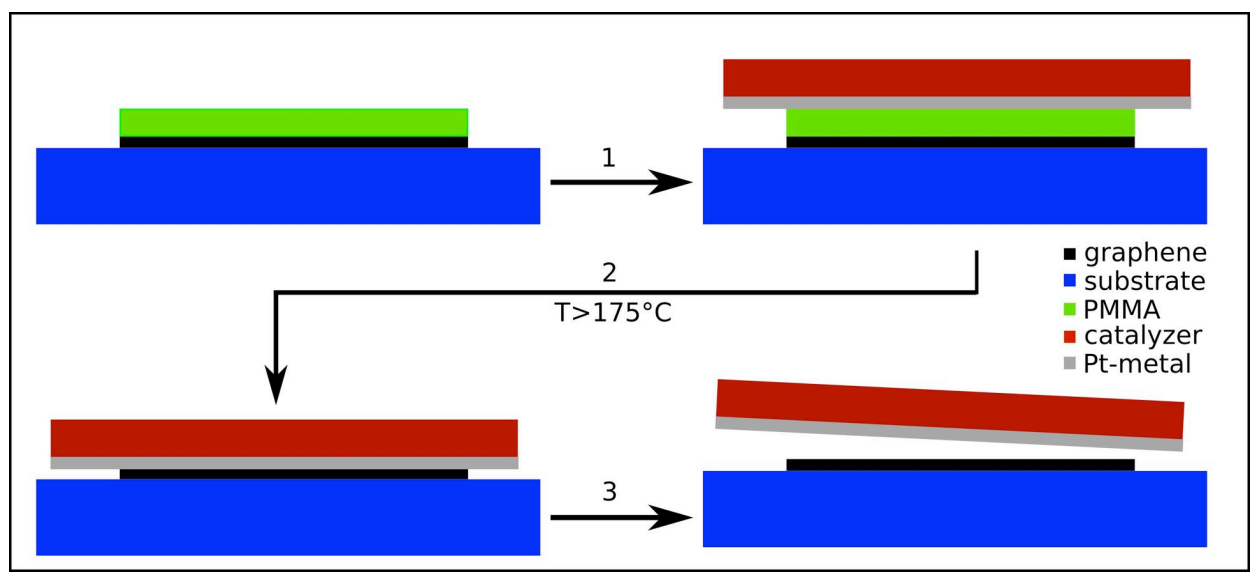

*Figure 1* Flow chart for interfacing ultraclean graphene with an arbitrary substrate structure.

**MATERIALS AND METHODS**

**Transfer**

All graphene layers used in this study are CVD grown on polycrystalline copper and a 950K PMMA capping (about 200 nm thick) is spin-coated onto the graphene. The preparation of clean graphene starts with chemical wet etching of the copper substrate by a 20% FeCl$_3$ solution. After completion of the etching process the PMMA/graphene composite is rinsed four times floating on uhpH$_2$O. Thereafter, it is transferred onto any desired device, here a SiO$_2$/Si wafer, and undergoes a first soft annealing process for 10 min at 75°C to remove any excess water. As illustrated in Figure 1, a platinum metal (Pt or Pd)-coated structure is then brought in close proximity with the PMMA capping and the entire unit is subsequently simply annealed in air at a temperature between 175 and 350 °C; i.e. well below the oxidation temperature of graphene [11,19]. Finally, the catalyst is removed, revealing ultraclean graphene interfaced with the desired structure.

**Low-energy electron transmission images**

The low-energy electron transmission investigations presented below are recorded in a low-energy electron holography setup inspired by the original idea of Gabor[20,21] of in-line holography where a sharp (111)-oriented tungsten tip acts as source of a divergent beam of highly coherent electrons[15,17,22]. The electron emitter can be brought as close as 200 nm to the sample with the help of a 3-axis piezo-manipulator. Part of the electron wave impinging onto the sample is elastically scattered and represents the object wave, while the un-scattered part of the wave represents the reference wave. At a distant detector, the hologram, i.e. the interference pattern between the object wave and the reference wave is recorded. The magnification in the image is given by the ratio of detector-tip-distance to sample-tip-distance and can be as high as one million.

**RESULTS AND DISCUSSION**

Figure 2, illustrates the procedure described above using the example of two partially overlapping graphene sheets interfaced with an oxidized silicon chip. Here, the oxide layer (90nm) is just needed to provide the interference contrast to identify individual graphene sheets[23]. For applications aiming at mesoscopic integrated electronic devices, such oxide layer will provide the needed electrical insulation of the graphene sheet for employing the two-dimensional electron gas in single- or multi-layer graphene.

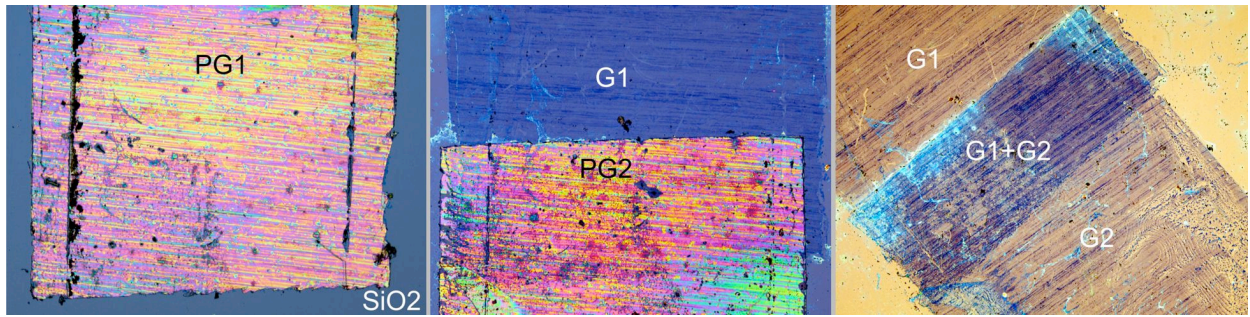

*Figure 2. Optical microscopy images of the subsequent deposition of two graphene sheets onto an oxidized Si-chip. Left: a polymer graphene composite (PG1) has been transferred onto the chip with the PMMA capping layer facing up. Middle: after catalytic removal of the PMMA, the bare graphene layer (G1) remains on the Si-chip and subsequently a second polymer graphene composite (PG2) has been deposited as to overlap the first layer (G1). Right: After applying the platinum metal catalysis a second time, the two graphene layers (G1 and G2) are apparent as well as the region where they overlap (G1+G2). The field of view in each image corresponds to about 3 mm$^2$.*

Figure 3 illustrates another example of interfacing graphene with an oxidized Si-chip of 10x10 mm² in size. Again, the 90 nm thick oxide layer provides the interference contrast that allows to distinguish between single, double and triple layers once the process described above has been repeated three times in such a way that the three macroscopic graphene layers partially overlap.

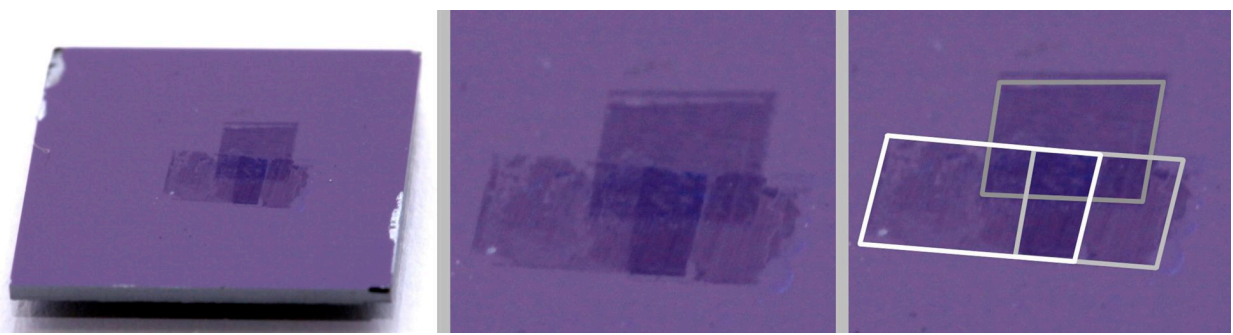

*Figure 3.* Optical photograph of three overlapping graphene sheets subsequently deposited onto a 10x10 mm² oxidized Si-chip and catalytically cleaned. The 90 nm thick native oxide layer provides the contrast between single, double and triple layers of graphene.

In order to examine the level of cleanliness of the graphene and compare it to the one of conventionally cleaned graphene by means of low-energy electron transmission microscopy, the samples have been deposited over an array of 2 micron holes in a titanium-gold covered SiN membrane. Low-energy electron transmission images were recorded using a setup for low-energy electron holography as reported elsewhere[17,24,25]. The use of electrons in the kinetic energy range of 50-250 eV implies high sensitivity for detecting possible residues on the freestanding graphene.

As evident from Figure 4, the method introduced here removes PMMA much more rigorously than the conventional procedure. In fact, the wet etched sample is essentially opaque for low-energy electrons, transmission can only be detected if excessive currents and exposure times are applied. In order to display the PMMA contamination, as shown on the right in Figure 4, twice the primary electron current and an exposure four times as long were needed to acquire the projection image.

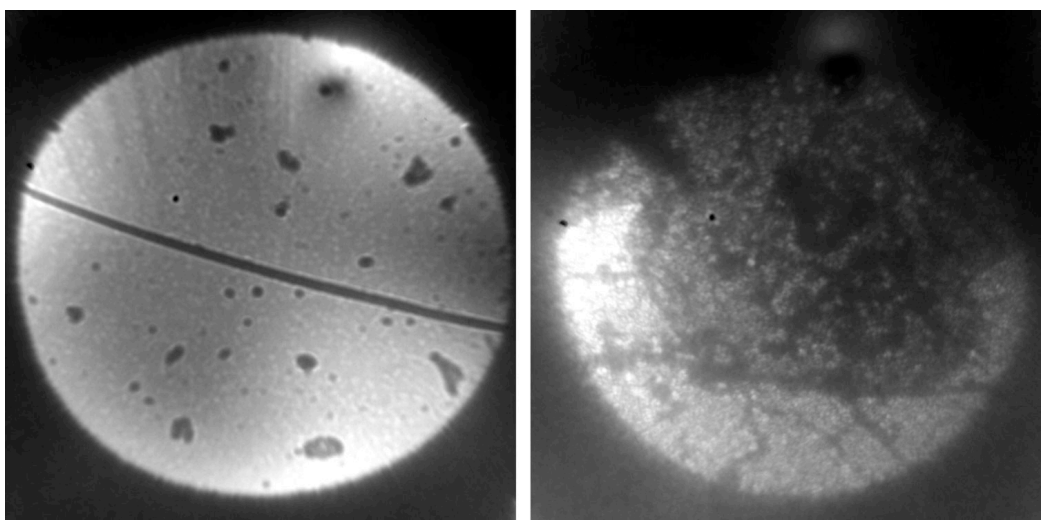

*Figure 4.* *Comparison between catalytically cleaned graphene (left) and conventionally prepared graphene (right), inspected by the transmission of 70 eV electrons. Transmission through conventionally cleaned graphene is hardly detectable: that is why the image on the right has been acquired with twice the electron current and an exposure time four times as long. Both samples experienced the same thermal treatment prior to the insertion into the vacuum system.*

Of course, the structural quality of the transferred graphene is given by the CVD process of growing graphene on polycrystalline copper. Figure 5 shows that the commercially available samples used here are not single layer graphene over large distances but contain also double as well as triple layers. However, these multilayers if subject to the catalytic process appear exceptionally clean as well.

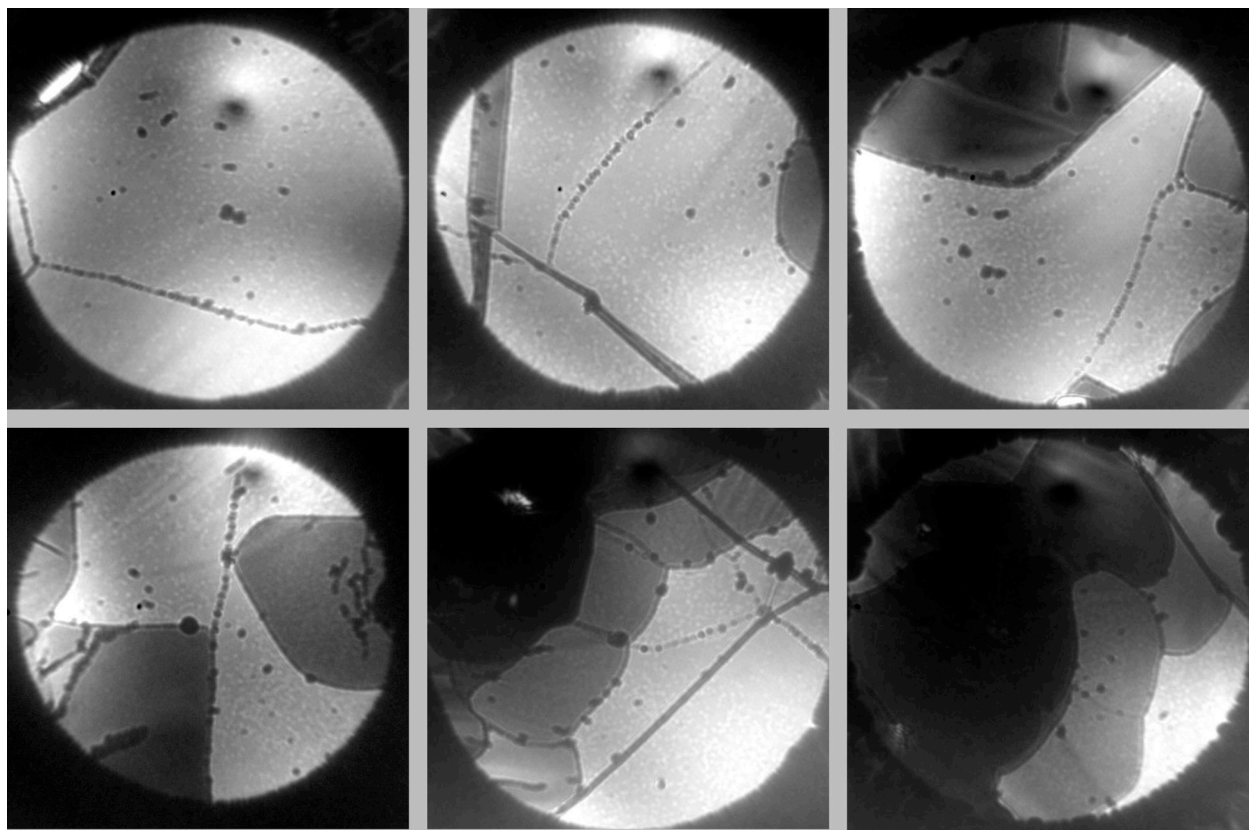

*Figure 5*. *Examples of electron transmission through graphene transferred onto an array of 2 micron holes milled in a titanium-gold covered SiN membrane. It is apparent that the commercially available graphene we used exhibits single, double and triple layers which are attributed to the CVD growth process on polycrystalline copper.*

The thermal degradation of polymers is a complicated process and subject of an entire research field[26,27]. Various decomposition mechanisms have been proposed, such as random-chain scission, end-chain initiation, unzipping, depropagation or depolymerization, to name just a few. For PMMA in particular, the end-chain initiation has been indentified as the predominant process[26]. A common aspect in all the different decomposition paths is the involvement of atomic hydrogen. Normally, during the thermal degradation of polymers, hydrogen originates from either the backbone or the side chains of the macromolecule, resulting in the formation of smaller molecules or the degradation into monomers. However, this process requires high temperatures, at least 400 °C for PMMA[26]. If the fragments produced are small enough, their low vapor pressure leads to desorption into the gas phase. In the case presented here, the catalytic aspect of the process can possibly be described as hydrogenation of the PMMA promoted by the platinum metals such that the cracking of the polymer occurs already at much lower temperatures. Platinum metals are well known in catalysis and it is conceivable that the decomposition is initiated by the ability of the Pt-metal to dissociate molecular adsorbed $H_2$ into atomic hydrogen. The fact that a minimal temperature must be attained to start such reaction (175 °C in our case), that the time for completing the reaction decreases rapidly with increasing temperature, and that with other metals such as gold the reaction is not actuated at all, are additional strong indications for the catalytic character of the process described here. In fact, preliminary experiments show that the time needed to remove a PMMA layer of well-defined thickness versus the applied temperature follows an Arrhenius behavior. This suggests that the thermally activated catalytic reaction is associated with an activation barrier.

**CONCLUSION**

In summary, we have demonstrated that the use of platinum metals for the transfer of graphene leads to ultraclean layers. The decomposition process of the PMMA layer is of catalytic nature and is promoted by the presence of a platinum metal. Already in its proximity the polymer layer is removed completely, revealing clean graphene on an arbitrary substrate or even freestanding. The degradation reaction proceeds in air and at temperatures ranging from 175 to 350 °C. This preparation method is thus easily applicable in every laboratory and does not require any special equipment. With this, ultraclean graphene is now routinely available to serve in several applications such as novel mechanical or electronic mesoscopic devices or as molecular sieves where ultraclean graphene is needed as a pre-requisite.

**ACKNOWLEDGMENT**

The authors are grateful for financial support by the Swiss National Science Foundation.